\documentclass[12pt]{article}
\usepackage[margin=1in]{geometry}

\usepackage{amssymb,amsmath,natbib,graphicx,changepage,relsize,amsthm,enumerate,bbm,subcaption,url,tikz}
\usepackage[hidelinks]{hyperref}

\theoremstyle{plain}
\newtheorem{theorem}{Theorem}
\newtheorem{theorem1}{Theorem}

\newtheorem*{lemma*}{Lemma}

\newtheorem{result}[theorem]{Result}
\newtheorem{algorithm}[theorem1]{Algorithm}

\theoremstyle{remark}

\newcommand{\ones}{\mathbbm{1}}

\newcommand{\cov}{\mbox{Cov}}

\newcommand{\const}{\mbox{const}}
\newcommand{\xtilde}{\widetilde{X}}

\newcommand{\Daxb}{D_{\mathsmaller{AX\beta}}}

\newcommand{\bmu}{\boldsymbol\mu}
\newcommand{\bnu}{\boldsymbol\nu}

\newcommand{\by}{\boldsymbol{y}}

\newcommand{\bbeta}{\boldsymbol{\beta}}

\newcommand{\beps}{\boldsymbol\epsilon}
\newcommand{\bgamma}{\boldsymbol\gamma}

\newcommand{\diag}{\mbox{{\bf diag}}}
\newcommand{\Diag}{\mbox{Diag}}

\begin{document}

\begin{center}
{\Large\bf Heterogeneous Susceptibilities in Social Influence Models}

\vspace{0.5in}

Daniel K. Sewell
\footnote{Daniel K. Sewell is Assistant Professor, Department of Biostatistics,
University of Iowa, Iowa City, IA 52242 (E-mail: {\it daniel-sewell@uiowa.edu}).}
\end{center}

\begin{abstract}
Network autocorrelation models are widely used to evaluate the impact of social influence on some variable of interest.  This is a large class of models that parsimoniously accounts for how one's neighbors influence one's own behaviors or opinions by incorporating the network adjacency matrix into the joint distribution of the data.  These models assume homogeneous susceptibility to social influence, however, which may be a strong assumption in many contexts.  This paper proposes a hierarchical model that allows the influence parameter to be a function of individual attributes and/or of local network topological features.  We derive an approximation of the posterior distribution in a general framework that is applicable to the Durbin, network effects, network disturbances, or network moving average autocorrelation models.  The proposed approach can also be applied to investigating determinants of social influence in the context of egocentric network data.  We apply our method to a data set collected via mobile phones in which we determine the effect of social influence on physical activity levels, as well as classroom data in which we investigate peer influence on student defiance.  With this last data set, we also investigate the performance of the proposed egocentric network model.

\vspace{ 2mm}

\end{abstract}

\section{Introduction}
Social influence can help explain behaviors, opinions and beliefs, and as such is of importance in sociology, marketing, public health, political science, etc.  \cite{newcomb1951social} describes social influence thus:
\begin{quote}
Any observable behavior [e.g., a displayed position on an issue] is not only a response (on the part of a subject) which is to be treated as a dependent variable;  it is also a stimulus to be perceived by others with whom the subject interacts, and thus to be treated as an independent variable.
\end{quote}
A prodigious amount of research has been done evaluating the effects of social influence on various attributes.  Examples include social influence on binge eating \citep{crandall1988social}, smoking and drinking behaviors of youth \citep{unger2001peer}, investment decisions \citep{hoffmann2009susceptibility}, emotions \citep{hareli2008emotion}, and transitioning from noninjecting heroine user to an injecting user \citep{neaigus2006transitions}.

The statistical models of choice for accounting for and evaluating the impact of social influence has long been the class of network autocorrelation models.  These have been deemed the `workhorse for modeling network influences on individual behavior' \citep{fujimoto2011network}.  These models have their roots in spatial statistics, with important early works by \cite{ord1975estimation} and \cite{doreian1980linear}.  These same models were quickly and successfully used for network data, and are still widely used and studied.

Network autocorrelation models make the strong assumption of homogeneous susceptibilities to social influence among network members; that is, each actor in the network is assumed to be equally susceptible to peer influence.  This is contrary to much of the research being done by substantive scientists.  \cite{friedkin1999social} 
devised a theory of social influence that incorporates heterogeneity in susceptibility to influences through the network.  Empirical studies have also shown that susceptibility can in fact vary based on subject attributes.  For example, studies have shown that birth order \citep{staples1961anxiety}, age \citep{krosnick1989aging,steinberg2007age}, and gender \citep{eagly1978sex} all affect an individual's resistance to social influence;  \cite{fennis2012revisiting} show that reducing personal control leads to a higher susceptibility to social influence;  \cite{urberg2003two} show that relationship variables can increase conformity to one's peers with respect to substance-use in adolescents; countless papers have been published using the Consumer Susceptibility to Interpersonal Influence measure \citep{bearden1989measurement}, determining a subject's susceptibility to social influence.
%

Using network autocorrelation models as a starting point, we relax the assumption of uniform susceptibility to social influence.  This is done by using a non-linear hierarchical model in which an individual's susceptibility is a function of that individual's characteristics.  These characteristics can take on a variety of forms; the most obvious of these is an individual's attributes such as gender, but one may also use local network topological features, such as centrality measures, and in so doing determine how the network topology itself can affect how an actor may be susceptible to social influence.  The proposed model can follow from the major network autocorrelation models, namely the Durbin model, the network effects model, the network disturbances model, and the network moving average model.  Just as in homogeneous network autocorrelation models, the posterior distribution for our model is available in closed form but is not a well-known distribution.  However, by using the Laplace approximation, we show how to very quickly obtain approximate samples from the posterior.  

Section \ref{methods} describes our proposed methods and our approximation to the posterior distribution.  In Section \ref{egocentricNetworks}, we extend these ideas to egocentric network data.  Section \ref{simulationStudy} describes a simulation study.  Section \ref{healthActivity} describes an analysis of data collected via mobile phones measuring call logs among network members and health activity.  Section \ref{peerInfluenceInTheClassroom} describes an analysis of educational data collected via surveys measuring student attitudes and behaviors.  We end with a brief conclusion in Section \ref{conclusion}.

\section{Methods}
\label{methods}
In this section, we first provide a brief review of the four most common types of network autocorrelation models.  We then describe the proposed model and Bayesian estimation procedure for the simplest of these, which assumes a simple diagonal covariance matrix for the response vector.  Next, we describe more sophisticated and realistic social influence models and how to adapt the estimation scheme accordingly.  We end the section by relating our work to a common operationalization of the adjacency matrix, namely row normalization.

Before beginning with the review, we shall provide some notation.  We will let $\by$ be the $n\times 1$ response vector of interest, where $n$ is the total number of actors in the network.  Let $X_1$ and $X_2$ be design matrices corresponding to the independent variables, having dimension $n\times p_1$ and $n\times p_2$ respectively.  Let $A$ be the $n\times n$ adjacency matrix such that $A_{ij}=1$ if there is an edge from actor $i$ to actor $j$ and zero otherwise;  $A_{ii}=0$ for all $i$.  In some cases $A$ may represent a weighted network where the non-diagonal entries of $A$ may take values in some subset of $\Re$; these values typically represent the strength of the edges in some meaningful way, e.g., the count of interactions between $i$ and $j$.  Let $W_x$ and $W_\epsilon$ be design matrices with dimension $n\times q_1$ and $n\times q_2$ respectively constructed from independent variables, functions of local network topologies, or some combination of the two.  Note that $X_1$, $X_2$, $W_x$, and $W_\epsilon$ may be identical, share some covariates, or be constructed from entirely different covariates.  
We will use $\diag(A)$ to represent the column vector constructed from the diagonal entries of some square matrix $A$, and $\Diag({\bf a})$ to represent the diagonal matrix constructed by setting the diagonal elements to be the elements of the vector ${\bf a}$.  Sometimes it will be necessary to refer to a row, column, or element of a matrix;  for any matrix $M$ this will be denoted by $M_{(i,\cdot)}$, $M_{(\cdot,j)}$, or $M_{(i,j)}$ respectively.

\subsection{Review of network autocorrelation models}
\label{Review}
The class of network autocorrelation models provide a solid statistical framework with which to investigate the effects of social influence, or, should social influence be considered a nuisance parameter, appropriately account for the complex dependencies in the data due to the network effect.  This class is typically associated with four statistical models.  The simplest of these is the Durbin model.  This model assumes that the observations are independent given the network and the covariates, but that an individual's mean is affected by the covariates of his/her neighbors.  Specifically, the Durbin model is given by
\begin{align}
\by&=X_1\bbeta_1 + \rho_x AX_2\bbeta_2 + \beps,
\label{DurbinEq}
\end{align}
where $\bbeta_1$ and $\bbeta_2$ are parameter vectors of unknown coefficients of size $p_1$ and $p_2$ respectively, $\rho_x$ is the parameter that captures the (uniform) social influence effect, and $\beps$ is a vector of independent mean zero normal random variables with variance $\sigma^2$.  Note that $\rho_x$ is constrained to equal 1 for model identifiability.  The assumption of independence is most often unreasonable in the context of network data.  To address this, the Durbin model can be augmented to allow for correlated errors.  Three ways to do this are the effects, disturbances, and moving average models \citep[see, e.g.,][]{doreian1980linear,hepple1995bayesian}.  

The network effects model is given by
\begin{align}
\by&=\rho_\epsilon A\by + X_1\bbeta_1 + \rho_xAX_2\bbeta_2 + \beps.
\label{effectsEq}
\end{align}
In this model, in addition to the effect of neighbors' covariates, an individual's mean response is a function of his/her neighbors' responses.  The network disturbances model is given by
\begin{align}\nonumber
\by&=X_1\bbeta_1 + \rho_xAX_2\bbeta_2 + \bnu,&\\
\bnu&=\rho_\epsilon A\bnu + \beps.&
\label{disturbancesEq}
\end{align}
Hence the network disturbances model is the Durbin model with the network effects model (sans covariates) on the errors.  This model can be interpreted to mean that an individual's deviation from his/her mean is a function of his/her neighbors' deviations from their mean.  Similar in spirit is the network moving average model, given by
\begin{align}
\by&=X_1\bbeta_1 + \rho_xAX_2\bbeta_2 + \beps + \rho_\epsilon A\beps.&
\label{MAEq}
\end{align}
The errors are additive based on the network structure, and hence an individual's response depends on the random fluctuations of his/her neighbors that cannot be explained by the neighbors' covariates.

\subsection{Durbin model}
\label{exogenous}
Here we relax the assumption of homogeneous susceptibility to social influence, beginning with the Durbin model.  If an individual's susceptibility is unique to them, then we may rewrite (\ref{DurbinEq}) as
\begin{align}\nonumber
\by&=X_1\bbeta_1 + R_xAX_2\bbeta_2+\beps,&
\end{align}
where $R_x$ is some diagonal matrix.  If we further hypothesize that an individual's susceptibility is determined by some set of covariates or local network functions contained in the design matrix $W_x$, we may let
\begin{align}
\diag(R_x)&=\widetilde W_x\widetilde\bgamma_x,&\\
\widetilde W_x&= (\ones, W_x),&\label{constraint1} \\ 
\widetilde\bgamma_x&=(1,\bgamma_x')',&\label{constraint2}
\end{align}
where $\ones$ is the vector of 1's.  The constraints given in (\ref{constraint1}) and (\ref{constraint2}) ensure that the model is identifiable, under the (obvious) assumptions that no columns of $W_x$ are proportional to $\ones$ and all design matrices are of full rank. 

We assume the priors on the parameters are of the following form:
\begin{align}
\bbeta&\sim N({\bf 0},g_1\sigma^2I_{p_1+p_2}),&\\
\bgamma_x&\sim N({\bf 0},g_2\sigma^2I_{q_1}),&\\
\sigma^2&\sim IG(a/2,b/2),&
\end{align}
where $\bbeta=(\bbeta_1',\bbeta_2')'$, $N(\bmu,\Sigma)$ is the multivariate normal distribution with mean $\bmu$ and covariance matrix $\Sigma$, and $IG(a,b)$ is the inverse gamma distribution with shape parameter $a$ and scale parameter $b$.  The log of the posterior distribution is
\begin{align}\nonumber
\ell&:=\log\big(\pi(\bbeta,\bgamma_x,\sigma^2|\by)\big)&\\
&=\const -\frac{a+n+p_1+p_2+q_1+2}{2}\log(\sigma^2) -\frac{1}{2\sigma^2}\left(b+
\|\by-\xtilde\bbeta\|^2 + \frac1{g_1}\|\bbeta\|^2+\frac1{g_2}\|\bgamma_x\|^2
\right),&
\label{logPostDurbin}
\end{align}
where $\xtilde=\begin{pmatrix}X_1 &R_xAX_2\end{pmatrix}$.  

For estimation, we propose using the Laplace approximation.  To this end, it is important to note the following facts, which can easily be confirmed:  For two $n\times1$ vectors ${\bf a}$ and ${\bf b}$ and any $n\times n$ matrix $M$, we have that
\begin{description}
\item[Fact 1.] $\Diag({\bf a}){\bf b}=\Diag({\bf b}){\bf a}$,
\item[Fact 2.] $\big[M'\Diag^2({\bf a})M\big]_{ij} = {\bf a}'\Diag(M_{(\cdot,i)})\Diag(M_{(\cdot,j)}){\bf a}$.
\end{description}
\noindent Thus using Fact 1 we can rewrite $R_xAX_2\bbeta_2$ as
\begin{equation}
R_xAX_2\bbeta_2 = (I_n + \Diag(W_x\bgamma_x))AX_2\bbeta_2 = AX_2\bbeta_2 + \Daxb W_x\bgamma_x,
\end{equation}
where $\Daxb :=\Diag(AX_2\bbeta_2)$.  The first derivatives can then be found to be
\begin{align} 
\frac{\partial \ell}{\partial\bbeta}&=-\frac1{\sigma^2}\left(
\left(\xtilde'\xtilde +\frac1{g_1}I_{p_1+p_2} \right)\bbeta - \xtilde'\by
\right),&\\ 
\frac{\partial\ell}{\partial\bgamma_x}&=-\frac1{\sigma^2}\left(
\left(W_X'\Daxb ^2W_X + \frac1{g_2}I_{q_1}\right)\bgamma_x - W_x'\Daxb (\by-X_1\bbeta_1-AX_2\bbeta_2)
\right),& \label{partialGamX}\\
\frac{\partial \ell}{\partial\sigma^2}&=-\frac{a+p_1+p_2+q_1+n+2}{2\sigma^2}+ \frac{1}{2(\sigma^2)^2}\left(
b + \|\by-\xtilde\bbeta\|^2 + \frac 1{g_1} \|\bbeta\|^2 + \frac1{g_2}\|\bgamma_x\|^2
\right).&
\end{align}

This then allows us to implement the following iterative algorithm to compute the maximum a posteriori (MAP) estimators.
\begin{algorithm}\label{algo1}
~\\ \vspace{-1pc}
\begin{description}
\item[0] Initialize $\widehat\bbeta$ and $\widehat\bgamma_x$.
\item[1] Set $R_x = \Diag(\widetilde W_x(1,\widehat\bgamma_x')')$.
\item[2] Set $\xtilde=\begin{pmatrix}X_1 &R_xAX_2\end{pmatrix}$.
\item[3] Set $\widehat\bbeta = \left(\xtilde'\xtilde +\frac1{g_1}I_{p_1+p_2} \right)^{-1}\xtilde'\by$.
\item[4] Set $\Daxb=\Diag(AX_2\widehat\bbeta)$.
\item[5] Set $\widehat\bgamma_x =\left(W_x'D_{\mathsmaller{AX\beta}} ^2W_x + \frac1{g_2}I_{q_1}\right)^{-1}W_x'D_{\mathsmaller{AX\beta}} \left(
\by-X_1\widehat\bbeta_1-AX_2\widehat\bbeta_2
\right)$.
\item Repeat steps 1 through 5 until reaching convergence.
\item[6] Set $\widehat{\sigma^2} = \frac1{a+p_1+p_2+q_1+n+2}\left(
b + \|\by-\xtilde\widehat\bbeta\|^2 + \frac 1{g_1} \|\widehat\bbeta\|^2 + \frac1{g_2}\|\widehat\bgamma_x\|^2
\right)$.
\end{description}
\end{algorithm}

Most of the second derivatives to construct the Jacobian are straightforward.  However, $\partial^2\ell/\partial\bgamma_x\partial\bbeta'$ is not entirely trivial.  First, consider just the $k^{th}$ element of (\ref{partialGamX}).  From Fact 2, we have
\begin{align*}
[W_x'\Daxb ^2W_x]_{k,\ell}& = \bbeta_2'X_2'A'\Diag(W_{x(\cdot,k)})\Diag(W_{x(\cdot,\ell)})AX_2\bbeta_2,&\\ \nonumber
\Rightarrow \frac{\partial\ell}{\partial\bgamma_{x(k)}}&=-\frac1{\sigma^2}\left(
\bbeta_2'X_2'A'\Diag(W_{x(\cdot,k)})(R_x-I)AX_2\bbeta_2 \right.& \\ 
& \left. - \bbeta_2'X_2'A'\Diag(W_{x(\cdot,k)})(\by-X_1\bbeta_1-AX_2\bbeta_2)
\right),& \\ \nonumber
\Rightarrow \frac{\partial^2\ell}{\partial\bbeta_2\bgamma_{x(k)}}&=\frac{1}{\sigma^2}X_2'A'\Diag(W_{x(\cdot,k)})\big(\by-X_1\bbeta_1-R_xAX_2\bbeta_2\big)& \\
&=\frac{1}{\sigma^2}X_2'A'\Diag\big(\by-X_1\bbeta_1-R_xAX_2\bbeta_2\big)W_{x(\cdot,k)},&
\end{align*}
and hence
\begin{align}
\frac{\partial^2\ell}{\partial\bbeta_2\partial\bgamma_x'}&=\frac{1}{\sigma^2}X_2'A\Diag(\by-X_1\bbeta_1-2R_xAX_2\bbeta_2)W_x.&
\end{align}

The above work yields the following result.
\begin{result}\label{result1}  The Laplace approximation to the posterior distribution corresponding to the Durbin model with heterogeneous susceptibilities to social influence is multivariate normal with mean equal to the MAP estimators $(\widehat\sigma^2,\widehat\bbeta,\widehat\bgamma)$, and covariance matrix equal to
\begin{equation*}
\cov\begin{pmatrix}
\sigma^2\\\bbeta \\ \bgamma_x
\end{pmatrix}=
\widehat{\sigma^2}\begin{pmatrix}
\frac{a+n+p_1+p_2+q_1}{2\widehat{\sigma^2}} &
{\bf 0} &
{\bf 0} \\
\cdot &
\widehat\xtilde'\widehat\xtilde+\frac1{g_1}I \hspace{1pc}&
\begin{pmatrix}
X_1'\widehat D_{\mathsmaller{AX\beta}} W_x \\
-X_2'A'\Diag(\by-X_1\widehat\bbeta_1-2\widehat R_xAX_2\widehat\bbeta_2)W_x
\end{pmatrix} \\
\cdot &
\cdot &
W_x'D_{\mathsmaller{AX\beta}}^2W_x + \frac1{g_2}I
\end{pmatrix}^{-1},
\end{equation*}
where $\widehat\xtilde$ and $\widehat D_{\mathsmaller{AX\beta}}$ are constructed using the MAP estimators.
\end{result}

A final note is that the Durbin model with heterogeneous susceptibilities is similar to a linear model with interaction terms between the $q_1$ variables from $W_x$ and $p_2$ variables obtained from $(AX_2)$, and in fact what we have proposed can be thought of as a subset of the interaction model.  In our proposed model, we are constraining the coefficient corresponding to the interaction between the $k^{th}$ variable in $W_x$ and the $\ell^{th}$ variable in $(AX_2)$ to have the form $\gamma_k\beta_{2\ell}$.  Obviously this has the potential benefit of parsimony by reducing the dimension of the parameter space (for the interaction terms) from $\Re^{p_2\cdot q_1}$ to $\Re^{p_2 + q_1}$, but much more importantly it allows us to interpret the coefficients in meaningful ways.

\subsection{Adding correlated errors}
\label{endogenous}
The previous section, while incorporating social influence through the covariates of an individual's alters, makes the strong assumption of independence of observations.  We now relax this assumption of independence while continuing to relax the assumption of homogeneous susceptibility to social influence.  To do this, we utilize the network effects, disturbances, and moving average models.  

As before, rather than having a single parameter $\rho_\epsilon$ representing the effect of social influence, we will let $\rho_\epsilon$ vary from individual to individual.  Specifically, we will use a diagonal matrix $R_\epsilon$ constructed by setting $\diag(R_\epsilon)=W_\epsilon\bgamma_\epsilon$.  We'll place a normal zero mean prior on $\bgamma_\epsilon$ with covariance $g_3\sigma^2I_{q_2}$.  After some manipulation, this yields the following result.

\begin{result}\label{result2}
The posterior distribution for network autocorrelation models with heterogeneous susceptibilities can be written in a general form:
\begin{align}\nonumber
\pi(\bbeta,\bgamma_x,\bgamma_\epsilon,\sigma^2|\by)
&\propto \left[|V|^{-\frac12}(b^*)^{-\frac{a^*}2}|\Sigma_\beta|^{\frac12}
\right]\cdot \left[
(b^*)^{\frac{a^*}2}(\sigma^2)^{-\frac{a^*}2-1}e^{-\frac{b^*}{2\sigma^2}}
\right] &\\
&\hspace{1pc}\cdot \left[
(\sigma^2)^{-\frac{p_1+p_2}{2}}|\Sigma_\beta|^{-\frac12}e^{-\frac1{2\sigma^2}(\bbeta-\bmu_\beta)'\Sigma_\beta^{-1}(\bbeta-\bmu_\beta)}
\right],
\label{generalPost}
\end{align}
where
\begin{align*}
a^*&=a+n+q_1+q_2,&\\
b^*&=b + \by'V^{-1}\by - \bmu_\beta'\Sigma_\beta^{-1}\bmu_\beta + \frac1{g_2}\|\bgamma_x\|^2 + \frac1{g_3}\|\bgamma_\epsilon\|^2, &\\
\Sigma_\beta^{-1}& = \xtilde'V^{-1}\xtilde + \frac1{g_1}I, &\\
\bmu_\beta &=\Sigma_\beta\xtilde'V^{-1}\by.&
\end{align*}
\begin{enumerate}[(a)]
\item The network effects model can be represented in this form by letting 
\begin{align*}
\xtilde&=(I-R_\epsilon A)^{-1}\begin{pmatrix}
X_1 &R_xAX_2
\end{pmatrix},&\\
V&=(I-R_\epsilon A)^{-1}(I-A'R_\epsilon)^{-1}.&
\end{align*}
\item The network disturbances model can be represented in this form by letting
\begin{align*}
\xtilde&=\begin{pmatrix}X_1 &R_xAX_2\end{pmatrix},&\\
V&=(I-R_\epsilon A)^{-1}(I-A'R_\epsilon)^{-1}.&
\end{align*}
\item The network moving average model can be represented by letting
\begin{align*}
\xtilde&=\begin{pmatrix}X_1 &R_xAX_2\end{pmatrix},&\\
V&=(I+R_\epsilon A)(I+A'R_\epsilon).&
\end{align*}
\end{enumerate}
\end{result}

There is no closed form solution to the MAP estimators iterative or otherwise, and so we cannot proceed directly as before.  However, with the grouping of terms we have used in (\ref{generalPost}), it is easy to see that $\bbeta$ and $\sigma^2$ can be integrated out of the posterior, leaving a closed form of $\pi(\bgamma_x,\bgamma_\epsilon|\by)$ equal (up to a multiplicative constant) to the first term in (\ref{generalPost}).  Therefore one may, using numerical algorithms, optimize this function and in so doing estimate the Hessian.  This then gives us a Normal approximation to that marginal posterior, which we will denote as $N(\bmu_\gamma,\Sigma_\gamma)$, where $\bmu_\gamma$ is the MAP of $\pi(\bgamma_x,\bgamma_\epsilon|\by)$, and $\Sigma_\gamma$ is the inverse of the negative Hessian matrix.  Given $\bgamma_x$ and $\bgamma_\epsilon$, we have that the conditional posterior distribution of $\bbeta$ and $\sigma^2$ is well known and easy to draw samples from.  Therefore, we may get approximate samples from the posterior distribution by the following algorithm.
\begin{algorithm}\label{algo2}
~\\ \vspace{-1pc}
\begin{description}
\item[0] Numerically find $\bmu_\gamma=\underset{(\bgamma_x'\hspace{0.3pc} \bgamma_\epsilon' )'}{\text{argmax}}\left\{|V|^{-\frac12}(b^*)^{-\frac{a^*}2}|\Sigma_\beta|^{\frac12}\right\}$ and set $\Sigma_\gamma=(-H)^{-1}$, where $H$ is the Hessian matrix.
\item For $m=1,\ldots,M$, repeat the following steps:
\item[1] Draw $\begin{pmatrix}{\bgamma_x^{(m)}}' & {\bgamma_\epsilon^{(m)}}' \end{pmatrix}'$ from $N(\bmu_\gamma,\Sigma_\gamma)$.
\item[2] Draw ${\sigma^2}^{(m)}$ from $IG\left(a^*/2, {b^*}^{(m)}/2)\right)$.
\item[3] Draw $\bbeta^{(m)}$ from $N(\bmu_\beta^{(m)},{\sigma^2}^{(m)}\Sigma_\beta^{(m)})$.
\end{description}
\end{algorithm}

\subsection{Extension of row normalization}
\label{rowNormalization}
\begin{figure}[h]
\centering
\begin{tikzpicture}[scale=2.5]
\node [circle,draw=black] (a) at (-1.42,0) {$ego_1$};
\node [circle,draw=black] (b) at (0,0) {$ego_2$};
\node [circle,draw=black] (c) at (-0.71,-0.71) {$alter_1$};
\node [circle,draw=black] (d) at (0.2,-0.98) {$alter_2$};
\node [circle,draw=black] (e) at (0.92,-0.38) {$alter_3$};
\node [circle,draw=black] (f) at (0.83,0.55) {$alter_4$};
\node [circle,draw=black] (g) at (0,1) {$alter_5$};
\draw [line width=1.5] (a) -- (c)
					  (b) -- (c)
					  (b) -- (d)
					  (b) -- (e)
					  (b) -- (f)
					  (b) -- (g);
\end{tikzpicture}
\caption{Toy pedagogical example for Section \ref{rowNormalization}.}
\label{toyRNExample}
\end{figure}
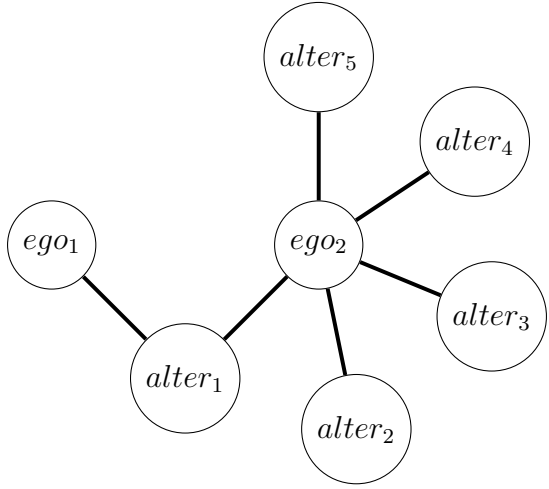

In the network autocorrelation model with homogeneous susceptibilities, there is a long tradition of row normalization \citep[e.g.,][or see \cite{leenders2002modeling} for a more general discussion on the operationalization of the adjacency matrix]{ord1975estimation,anselin1988spatial}.  The idea behind this is that every actor of the network receives the same amount of influence.  While this seems on a superficial level to enforce some degree of homogeneity among how actors are influenced, the opposite is in fact true.  Consider the toy example given in Figure \ref{toyRNExample}.  Ego 1 only has one tie, that with alter 1.  Ego 2 in contrast has 5 ties, that with alter 1 as well as four other ties.  Enforcing row normalization implies that ego 1 is much more susceptible to influence from alter 1 than is ego 2.  We can see, for example, that in the context of a moving average model, the covariance between ego 1 and alter 1 is $3\rho\sigma^2/2$, while the covariance between ego 2 and alter 1 is $7\rho\sigma^2/10$, even though both ego 1 and ego 2 are connected to alter 1 in the same way.  

Row normalization, in other words, assumes that how any two actors co-vary depends not only on whether or not the two actors are linked, but also by their respective local network topology, and disregards any attributes about the actors themselves.  This can also be seen by noting that row normalization is a special case of our proposed model, where $W_\epsilon$ is of rank 1 and equals the column vector consisting of the inverse of the local network sizes.  Using our proposed model we can now generalize this to allow the heterogeneity in susceptibilities to depend on multiple local topological features as well as covariates.

\section{Egocentric networks}
\label{egocentricNetworks}
Often times, collecting data on an entire network is infeasible.  Other times, the network boundaries are ill-defined, and so even given limitless resources it would not be possible to collect data on the entire network.  For these reasons, researchers often collect what is called egocentric network data.  That is, researchers select a subset of members of the network and collect measurements on these subjects and their alters, i.e., those individuals to whom there are ties with the original subset of network members.  There are some obvious disadvantages to this type of data, however, primarily because the dependence structure of the data depends on $A$, which is only partially observed.  \cite{sewell2017network} demonstrated how one may obtain inference on egocentric network data that is based on network autocorrelation models.  However, this approach continued to assume uniform susceptibility to social influence.  If the data generating process can be well approximated by a network moving average model, we can utilize our proposed approach when analyzing egocentric network data.  

To see this, first note that we can partition the data into non-overlapping subsets that correspond to the egos (i.e., the subset sampled), the egos' alters\footnote{Even though some egos may have edges involving other egos, `alters' here refer only to those egos' alters which have not been sampled.}, and all others.  Partitions associated with these three categories will be denoted by the subscripts $e$, $a$, and $o$ respectively.  Then (\ref{MAEq}) may be written as
\begin{align}\nonumber
\begin{pmatrix}
\by_e \\ \by_a \\ \by_o
\end{pmatrix}&=
\begin{pmatrix}
X_{1e}\\X_{1a}\\X_{1o}
\end{pmatrix}\bbeta_1 +
\begin{pmatrix}
R_{xe}&{\bf 0}&{\bf 0}\\
{\bf 0}&R_{xa} & {\bf 0}\\
{\bf 0}&{\bf 0}&R_{xo}
\end{pmatrix}
\begin{pmatrix}
A_e & A_{ea} &{\bf 0}\\
A_{ae} & A_a & A_{ao}\\
{\bf 0} & A_{oa} & A_o
\end{pmatrix}
\begin{pmatrix}
X_{2e}\\X_{2a}\\X_{2o}
\end{pmatrix}\bbeta_2 &\\
&+ 
\begin{pmatrix}
\beps_e\\ \beps_a \\ \beps_o
\end{pmatrix} +
\begin{pmatrix}
R_{\epsilon e}&{\bf 0}&{\bf 0}\\
{\bf 0}&R_{\epsilon a} & {\bf 0}\\
{\bf 0}&{\bf 0}&R_{\epsilon o}
\end{pmatrix}
\begin{pmatrix}
A_e & A_{ea} &{\bf 0}\\
A_{ae} & A_a & A_{ao}\\
{\bf 0} & A_{oa} & A_o
\end{pmatrix}
\begin{pmatrix}
\beps_e\\ \beps_a \\ \beps_o
\end{pmatrix}.&
\label{MAEgoEQ}
\end{align}
From (\ref{MAEgoEQ}) we can see that the distribution of the observed data $\by_e$ is
\begin{equation}
\by_e\sim N\Big(
X_{1e}\bbeta_1 + R_{xe}(A_eX_{2e}+A_{ea}X_{2a})\bbeta_2,
\sigma^2\big((I+R_{\epsilon e}A_e)(I+A_e'R_{\epsilon e}) + R_{\epsilon e}A_{ea}A_{ea}'R_{\epsilon e}
\big)\Big).
\end{equation}

With this setup, we have the following result.
\begin{result}\label{result3}
The network moving average model for egocentric data can be represented by (\ref{generalPost}), letting 
\begin{align*}
\xtilde & = \begin{pmatrix}
X_{1e} & R_{xe}(A_eX_{2e}+A_{ea}X_{2a}) \end{pmatrix}&\\
V& = (I+R_{\epsilon e}A_e)(I+A_e'R_{\epsilon e}) + R_{\epsilon e}A_{ea}A_{ea}'R_{\epsilon e},&
\end{align*}
and where $a^*$, $b^*$, and $\bmu_\beta$ are slightly changed to be
\begin{align*}
a^*&=a + n_e + q_1 + q_2,&\\
b^*&=b + \by_e'V^{-1}\by_e - \bmu_\beta'\Sigma_\beta^{-1}\bmu_\beta + \frac1{g_2}\|\bgamma_x\|^2 + \frac1{g_3}\|\bgamma_\epsilon\|^2, &\\
\bmu_\beta &=\Sigma_\beta\xtilde'V^{-1}\by_e.&
\end{align*}
\end{result}
\noindent This then lends itself to the same sampling procedure described in Algorithm \ref{algo2}.  

It should be possible to similarly combine the proposed methods for heterogeneous susceptibilities with the methods of \cite{sewell2017network} to implement either the network effects or disturbances model for egocentric network data.  
Doing so, however, dramatically increases the dimensionality of the estimation problem, and would require further work that is beyond the scope of this paper.  The Durbin model is, of course, trivially accommodated by fixing $R_{\epsilon}$ to be a matrix of zeros in Result \ref{result3}.

\section{Simulation study}
\label{simulationStudy}
We conducted a simulation study to examine the properties of the Bayesian estimators.  We simulated and analyzed 1,000 data sets and evaluated the Bayesian estimation by considering (1) the point estimation, (2) the credible interval coverage frequencies, and (3) the bias in estimating individual susceptibilities.

Each simulated data set was constructed in the following way.  To ensure that we were considering a realistic network topology, we used the network from Section \ref{healthActivity}.  We then generated a binary covariate by independent Bernoulli draws with success probability of $0.5$, and a continuous valued covariate by independent draws from a standard normal distribution.  These two variables constituted $X_2$ and, along with an intercept, $X_1$ (i.e., $X_1=\begin{pmatrix} \ones & X_2 \end{pmatrix}$).  The normally distributed covariate was also used in constructing $W_x$ and $W_\epsilon$, along with the inverse of the network size, betweenness, local clustering coefficient, and eigenvector centrality (and an intercept for $W_\epsilon$).  The network disturbances model was used in simulation and estimation.  We set all elements of $\gamma_x$ and $\gamma_\epsilon$ to equal 0.1, all elements of $\beta_1$ to equal 2, all elements of $\beta_2$ to equal 0.5, and we set $\sigma^2$ to equal 0.5.

\begin{figure}
\centering
\includegraphics[width=0.65\textwidth]{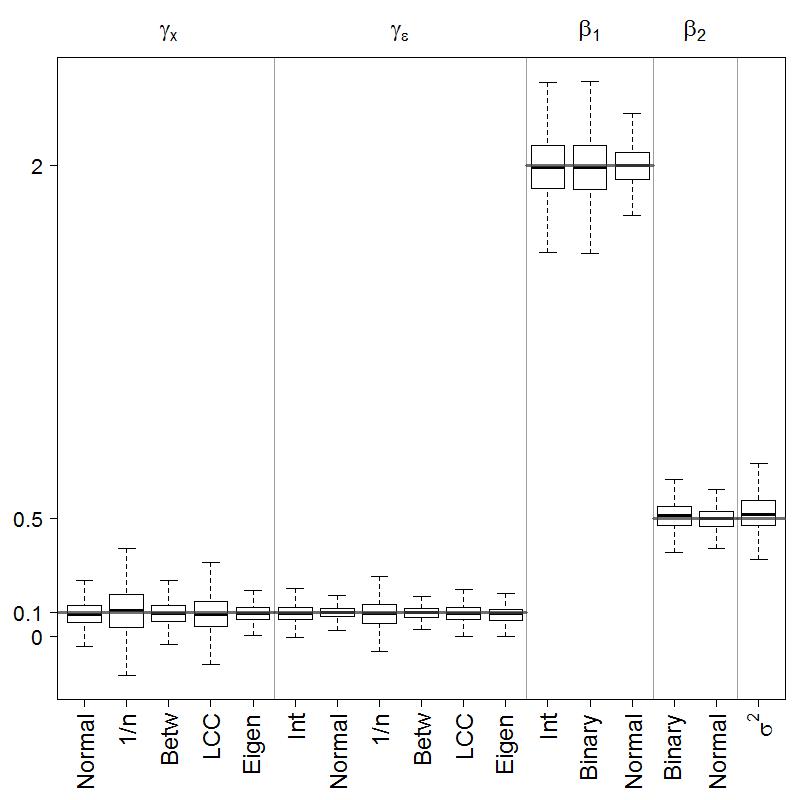}
\caption{Posterior means for the coefficients in $\bgamma_x$, $\bgamma_\epsilon$, $\bbeta_1$, and $\bbeta_2$ as well as the variance $\sigma^2$ corresponding to 1,000 simulated data sets.  Gray horizontal lines correspond to the true values.}
\label{simCoefsFig}
\end{figure}

\begin{figure}
\centering
\includegraphics[width=0.45\textwidth]{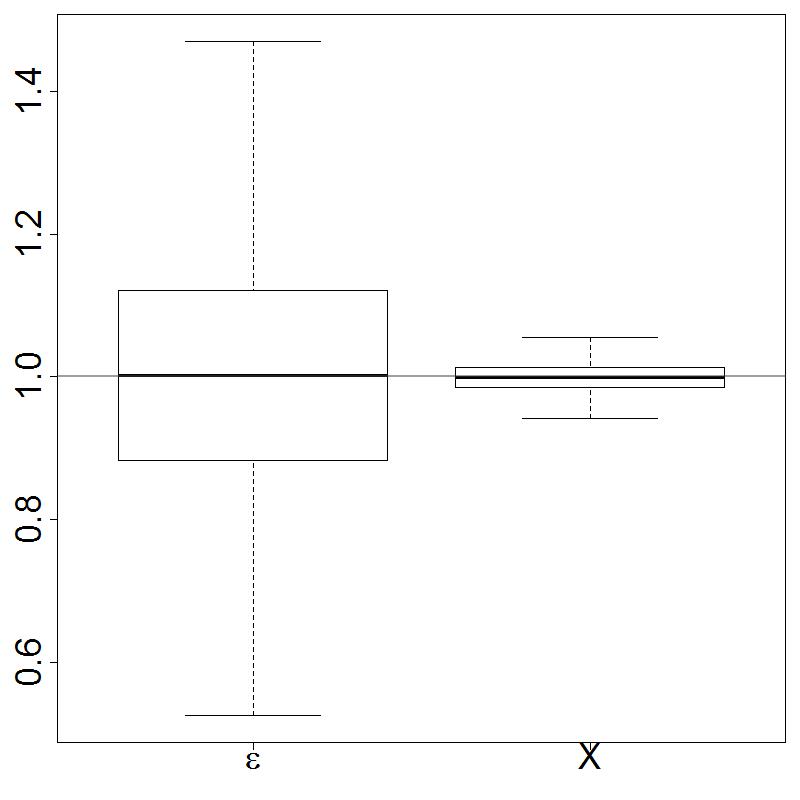}
\caption{Boxplots of median ratios of individual susceptibilities ($W_\epsilon\hat\gamma_\epsilon/W_\epsilon\gamma_\epsilon$ and $W_x\hat\gamma_x/W_x\gamma_x$) for the 1,000 simulations.}
\label{rhoBiases}
\end{figure}

Figure \ref{simCoefsFig} shows boxplots of the posterior means of the parameters.  The gray horizontal lines indicate the true value, and from this we confirm that the posterior means provide accurate point estimation of the parameters.  Of note is that there was non-negligible correlation between the local network measures (ranging from -0.61 to 0.42), but this did not appear to negatively affect the estimation performance.  Table \ref{simCovTab} provides the coverage frequencies of the 95\% credible intervals, i.e., the proportion of times the credible interval covered the true value.  Note that these are credible intervals and not confidence intervals, and so we do not necessarily expect the coverage frequency to be 95\% except in the limit as $n\rightarrow\infty$.  Nevertheless, with the sole exception of the eigenvector centrality term in $\gamma_\epsilon$, the coverage frequencies are close to 95\% with $n=122$.  Finally, network autocorrelation models have been shown to have a negative bias in estimating $\rho_\epsilon$ \citep[see, e.g.,][]{mizruchi2008effect,fujimoto2011network}.  Rather than evaluating bias in a single susceptibility parameter, our proposed models have $\mathcal{O}(n)$ susceptibilities to consider.  Thus to determine if there was any bias, for each simulated data set we computed the median of the ratios of $W_\epsilon\hat\gamma_\epsilon/W_\epsilon\gamma_\epsilon$ and $W_x\hat\gamma_x/W_x\gamma_x$, i.e., the ratios of estimated vs. true individual susceptibilities to social influence.  Figure \ref{rhoBiases} gives the boxplots of these medians for the 1,000 simulations.  From this we see that the proposed algorithm is doing a good job at estimating the susceptibilities of the individuals, though with more variability in estimating those susceptibilities associated with the disturbances than with the covariates; importantly (and perhaps surprisingly) we see no evidence of bias, negative or otherwise.

\begin{table}
\centering
\begin{tabular}{lr|lr|lr}
\hline 
Normal ($\bgamma_x$)& 0.932 & Intercept ($\bgamma_\epsilon$) & 0.895 & Intercept ($\bbeta_1$) & 0.942  \\ 
Inverse of network size ($\bgamma_x$)& 0.924 &Normal ($\bgamma_\epsilon$)& 0.901& Binary ($\bbeta_1$) & 0.948 \\ 
Betweenness ($\bgamma_x$)& 0.929 &  Inverse of network size ($\bgamma_\epsilon$)& 0.897  & Normal ($\bbeta_1$) & 0.962 \\ 
Loc. clust. coef. ($\bgamma_x$)& 0.933 &Betweenness ($\bgamma_\epsilon$)& 0.877 & Binary ($\bbeta_2$) & 0.922 \\ 
Eigenvector centrality ($\bgamma_x$)& 0.901 & Loc. clust. coef. ($\bgamma_\epsilon$)& 0.917& Normal ($\bbeta_2$)& 0.922 \\ 
 &&  Eigenvector centrality ($\bgamma_\epsilon$) & 0.708 & $\sigma^2$ & 0.882  \\ 
\hline
\end{tabular}
\caption{Coverage frequencies of 95\% credible intervals corresponding to 1,000 simulations.}
\label{simCovTab}
\end{table}

\section{Social influence on physical activity}
\label{healthActivity}
Much research in public health relates to the determinants of physical activity.  One of the most important such determinants is social support \citep{anderson2006social}.  Indeed, social influence has been found to be even more important than physical environmental factors \citep{giles2002relative}.  Implementing our proposed approach can be used to further our understanding in this area by learning what characteristics may affect the susceptibility to social influence with respect to engagement in physical activity.

We analyzed data collected from a study conducted on subjects in a young-family residential living community near a major research institution in North America \citep{aharony2011social}.  Data was collected via mobile phone technology developed by Aharony et al.  
The network constructed from this data that we studied consisted of 122 members.  This is a rich data set, and while there are many questions that could be addressed, we focus on evaluating the impact of social influence on physical activity levels.

Daily physical activity was measured by accelerometry data collected via the users' mobile phone device.  Each user was given an energy expenditure score by assigning points according to the amount of time of recorded physical activity.  See \cite{aharony2011social} for more details.  For each individual we averaged all daily values and used this as the response variable.  A call log was recorded for each subject, which allowed us to construct the adjacency matrix $A$, such that $A_{ij}=1$ if network members $i$ and $j$ called each other during the study period, indicating that the two members knew each other.  
We included as covariates the demographic variables gender and race, where race was categorized as Asian, Black, Hispanic, Middle Eastern, White, or other.  We also included the average monthly amount spent on discretionary purchases, average amount of sleep per night, and a self reported measure of quality of life.  This last measure was constructed in the following way.  Each of the network members was presented with the following sequence of statements with which individuals would agree or disagree:
\begin{enumerate}
\setlength\itemsep{0.1 pc}
\item In most ways my life is close to my ideal.
\item The conditions in my life are excellent.
\item I am satisfied with my life.
\item So far I have gotten the important things I want in life.
\item If I could live my life over, I would change almost nothing.
\item In general, I consider myself a happy person.
\item Compared to most of my peers, I consider myself more happy.
\item I am generally very happy. I enjoy life regardless of what is going on, getting the most out of everything.
\end{enumerate} 
Individuals responded to each statement using a seven point Likert scale, with larger values indicating stronger agreement.  We used as a covariate the average of the answers to all these questions; hence large values correspond to a better self reported quality of life.  
  
Of the covariates just described, all were included in $X_1$; $X_2$ was constructed from the variables reflecting choices or attitudes, i.e., the average monthly amount of discretionary purchases, quality of life, and amount of sleep.  We constructed $W_x$ by using gender, the inverse of network size, betweenness centrality, 
local clustering coefficient (LCC), and eigenvector centrality.  An actor $j$'s betweenness centrality is determined by considering the proportion of shortest paths through the network between pairs of actors that pass through $j$, and is often related to information flow.  An individual's LCC captures how densely connected his/her alters are and is strongly related to structural holes and control over information flow.  Eigenvector centrality extends the basic notion of degree by regarding not just the number of alters but also the importance of the alters to whom an individual is connected; thus, an individual with a large eigenvector centrality tends to be connected to many important individuals.  See, e.g., \cite{newman2010networks} more on these topics.  We then set $W_\epsilon=\begin{pmatrix}\ones & W_x\end{pmatrix}$.  All the non-categorical covariates, including the local network features, were standardized, and so the units of the original measurements are irrelevant.  

We implemented a network disturbances model, thus allowing an individual's physical activity to be influenced by the deviations of his/her alters' physical activity from what would be their expected activity levels.  That is, this model hypothesizes that if an ego's alter is doing more or less physical activity than what the ego would expect from that alter, the ego may be inclined to behave in a similar way.  We set $g_1=g_2=g_3=n(=122)$, $a=2.1$, and $b=70,000$.  We compared the proposed approach to performing the same Bayesian analysis except with homogeneous social influence, i.e., 
$\widetilde W_x=W_\epsilon=\ones$.  For both methods we drew 10,000 samples from the posterior distribution.  

\begin{figure}
\centering
\includegraphics[width=0.5\textwidth]{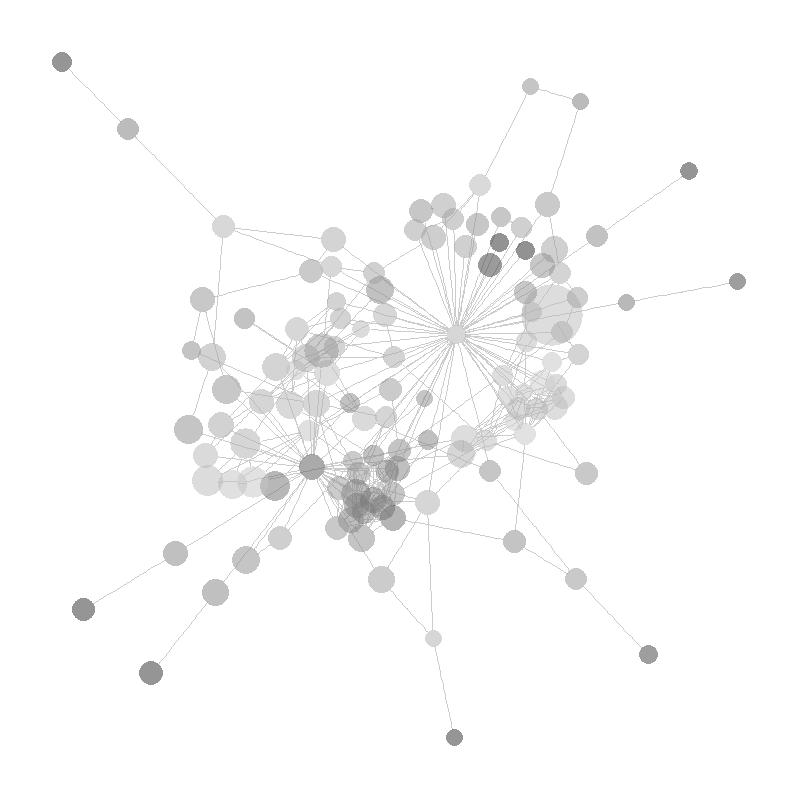}
\caption{Physical activity network.  Darker shading indicates higher susceptibility to social influence.  The size of the circles corresponds to the amount of physical activity.}
\label{FFNetFig}
\end{figure}

Figure \ref{FFNetFig} shows the network, where darker shading indicates higher susceptibility to social influence, and the size of the circle corresponds to the amount of physical activity.  The posterior means, standard errors, and 95\% credible intervals of the parameters are given in Table \ref{FFResults}.  From this we see that while the covariates of the alters do not seem to influence the egos, there is strong evidence to suggest that the alters' deviations from their expected activity levels do.  This analysis seems to provide some support to the idea of row normalization, as seen from the fact that most of the posterior probability mass is away from zero for the inverse of the local network size.  However, also of note is that eigenvector centrality seems to play a role as well;  those individuals with larger eigenvector centrality tended to be more susceptible to social influence.  Unsurprisingly, the estimate of the variance is larger for the model assuming homogeneous social influence.  This can be understood by recognizing that more of the individual level variability is explained by the varying degrees of susceptibility to the influence exerted by the alters' deviations.  One would expect the practical effect of this to be larger CI's, which is what we observed in this case.  In our data analysis, this played an important role with amount of sleep in $X_1$, as the 95\% credible interval only contained zero when heterogeneity in susceptibility was ignored.

\begin{figure}
\centering
\includegraphics[width=0.5\textwidth]{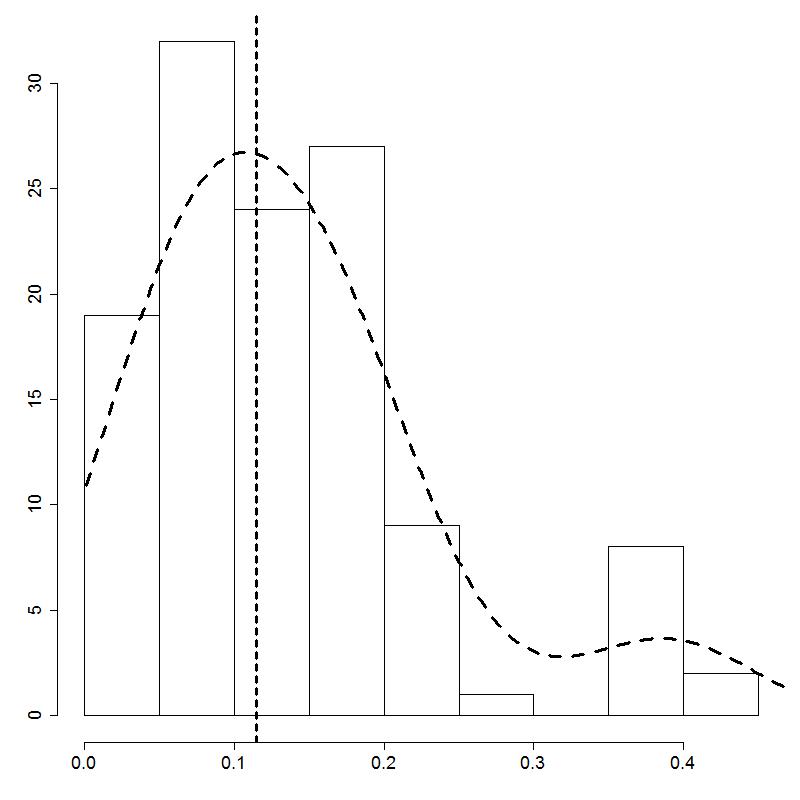}
\caption{Posterior means for individual susceptibilities to social influence through alters' deviations, i.e., $\diag(R_\epsilon)$.  The vertical dotted line gives the posterior mean for $\rho_\epsilon$ in the model ignoring heterogeneous susceptibilities.}
\label{rhoE}
\end{figure}

Figure \ref{rhoE} provides a histogram (and smoothed density estimates) of the posterior means of individual susceptibilities, i.e., the diagonal entries of $R_\epsilon$.  From this figure we corroborate that there is considerable non-negligible and heterogeneous susceptibility to social influence through the alters' deviations, and note that there exists a group of ten actors with particularly high susceptibility.

\begin{table}[h!]
\centering
\begin{tabular}{crrrrr}
&\multicolumn{2}{c}{Estimate} &&&\\
 & Hetero. & Homo.& SE & LB & UB \\ 
  \hline
Male ($\bgamma_x$) & -0.662 &  & 0.391 & -1.429 & 0.104 \\ 
Inverse of network size ($\bgamma_x$)& 0.551 &  & 0.437 & -0.306 & 1.408 \\ 
Betweenness ($\bgamma_x$)& -0.056 &  & 0.116 & -0.283 & 0.171 \\ 
Loc. clust. coef. ($\bgamma_x$)& 0.114 &  & 0.306 & -0.485 & 0.713 \\
Eigenvector centrality ($\bgamma_x$) & 0.522 &  & 0.277 & -0.021 & 1.066 \\ \hline
Intercept ($\bgamma_\epsilon$) & 0.115* & 0.114* & 0.042 & 0.032 & 0.197 \\ 
Male ($\bgamma_\epsilon$) & 0.045 &  & 0.046 & -0.045 & 0.136 \\ 
Inverse of network size ($\bgamma_\epsilon$) & 0.095* &  & 0.042 & 0.013 & 0.177 \\ 
Betweenness ($\bgamma_\epsilon$)& -0.010 &  & 0.008 & -0.025 & 0.006 \\ 
Loc. clust. coef. ($\bgamma_\epsilon$) & -0.025 &  & 0.024 & -0.072 & 0.023 \\ 
Eigenvector centrality ($\bgamma_\epsilon$) & 0.071* &  & 0.030 & 0.011 & 0.130 \\  \hline
Intercept ($\bbeta_1$) & 341.842* & 318.989* & 47.645 & 251.329 & 436.060 \\ 
Male ($\bbeta_1$) & -10.213 & -44.765 & 47.395 & -101.544 & 86.408 \\
Black ($\bbeta_1$)& -72.453 & 8.417 & 165.862 & -399.718 & 248.383 \\ 
Hispanic ($\bbeta_1$) & 273.359* & 278.066* & 90.201 & 97.130 & 453.960 \\ 
Middle Eastern ($\bbeta_1$)& 85.327 & 65.020 & 94.480 & -104.673 & 272.309 \\ 
White ($\bbeta_1$)& 77.488 & 101.006 & 59.909 & -37.195 & 199.335 \\ 
Other Race ($\bbeta_1$) & 165.429 & 198.790 & 117.166 & -60.752 & 397.057 \\ 
Discretionary Purchase ($\bbeta_1$) & 31.324 & 33.324 & 21.810 & -11.709 & 73.530 \\ 
Sleep ($\bbeta_1$) & -45.021* & -44.194 & 22.563 & -89.236 & -0.732 \\
Quality of life ($\bbeta_1$) & -43.450 & -40.740 & 22.486 & -87.109 & 0.577 \\  \hline
Discretionary Purchase ($\bbeta_2$) & 15.663 & 11.586 & 11.618 & -7.476 & 38.669 \\ 
Sleep ($\bbeta_2$) & 13.174 & 11.899 & 14.644 & -15.353 & 42.635 \\ 
Quality of Life ($\bbeta_2$) & 22.220 & 12.094 & 12.355 & -3.070 & 47.189 \\  \hline
$\sigma$ & 208 & 222 & 13 & 185  & 237 
\end{tabular}
\caption{Results for physical activity data.  The baseline race is Asian.  LB and UB correspond to the lower and upper bounds of the 95\% credible intervals respectively.  The posterior mean parameter estimates are given for both the model with heterogeneous and with homogeneous susceptibility to social influence.  SE, LB, and UB all correspond to the model accounting for heterogeneity in susceptibility.  *'s indicate that the 95\% CI excluded zero.}
\label{FFResults}
\end{table}

\section{Student defiance in the classroom}
\label{peerInfluenceInTheClassroom}
Student defiance is a significant barrier to educational success in schools.  Despite much research done on this topic \citep[e.g.,][]{gregory2008discipline,way2011school,hart2017teacher}, this endemic problem still persists and requires further study.  \cite{mcfarland2001student} conducted a study to investigate student resistance in the classroom, and both behavioral/attitude variables and friendship ties were recorded through a self-report survey.  We analyze this data to determine not only whether peer influence affects one's own classroom misbehavior, but also what makes an individual susceptible to such influence.

\begin{figure}
\centering
\includegraphics[width=0.5\textwidth]{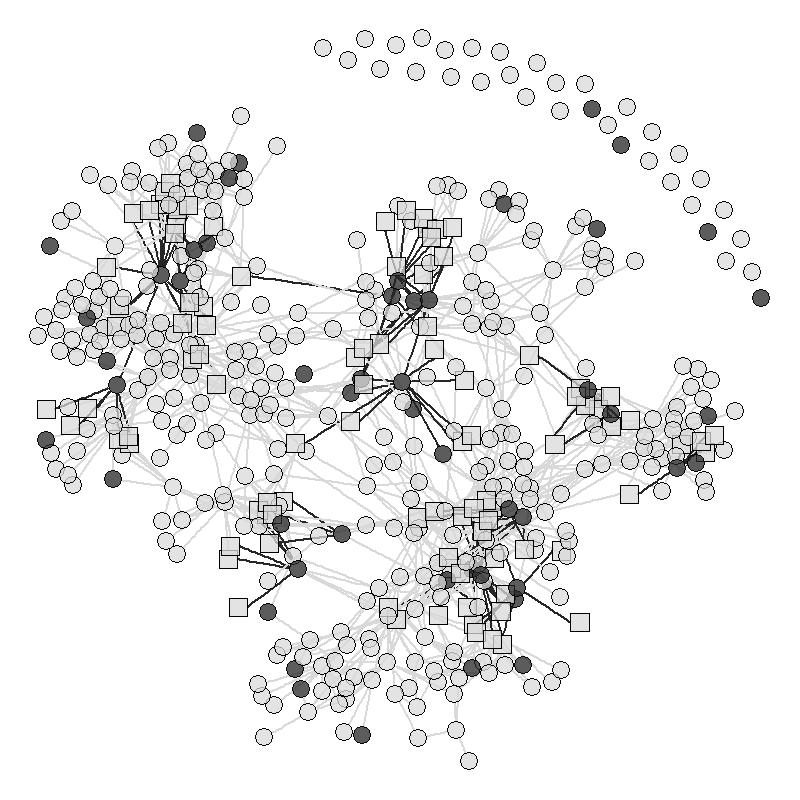}
\caption{Peer influence network.  The shading and shapes correspond to an example egocentric network, where the sampled egos are given by dark circles, their unsampled alters are given by gray squares, and their observed outgoing edges are shaded dark.}
\label{peerInflFig50}
\end{figure}

Our response variable is a self-reported frequency of misbehavior in class.  We included as covariates two Likert scale questions: 
\begin{enumerate}
\item (Q1.X) How important is this course for your future? 
\item (Q2.X) How much do you like the course subject?
\end{enumerate}
The responses range from 1 (not at all/strongly dislike) to 4 (very important/like it very much).  We also included the following Likert scale questions as covariates in the influence terms (i.e., in $W_\epsilon$):
\begin{enumerate}
\item (Q1.W) How interesting do you find your classmates?
\item (Q2.W) How often do you socialize?
\item (Q3.W) How interesting do you find your friends in this class? 
\end{enumerate}
The responses ranged from 1 (not at all/never/not at all) to 4 (very interesting/often/very interesting).  There were 472 complete observations used in our analysis.  Each student was asked to list their relationship to each other person in the class; we set $A_{ij}=1$ if student $i$ said they were friends or close friends with student $j$.  Figure \ref{peerInflFig50} shows this network.

We implemented a network moving average model with this data.  A primary goal was to investigate how quickly the results from an egocentric sample reflect those that one would obtain from the full data, and so we implemented not only the full network model of Section \ref{methods} but also the egocentric network model of Section \ref{egocentricNetworks}, where the number of egos $n_e$ ranged from 50 to 250 in increments of 50.  Figure \ref{peerInflFig50} shows an example of an egocentric network for 50 sampled egos.  For each sample size $n_e$ we drew 100 egocentric samples and analyzed the results.  For the full network analysis and for each of the $5\times100$ egocentric network analyses we used 10,000 posterior draws.

Table \ref{peerInflResultsTab1} provides the results from analyzing the full data.  We see that social influence is in fact playing a non-trivial role in student defiance and disruptive behaviors.  We also see that how often a student socializes affects that student's susceptibility to be influenced with respect to defiant behavior, specifically that this susceptibility decreases as the amount of socialization increases.  We did not include many local network features, as these often cannot be computed for egocentric network data; we did however include the inverse of the network size, though the posterior distribution seems to suggest that a row normalized term may not be pertinent.  

\begin{figure}
\centering
\begin{subfigure}[]{0.48\textwidth}
\includegraphics[width=\textwidth]{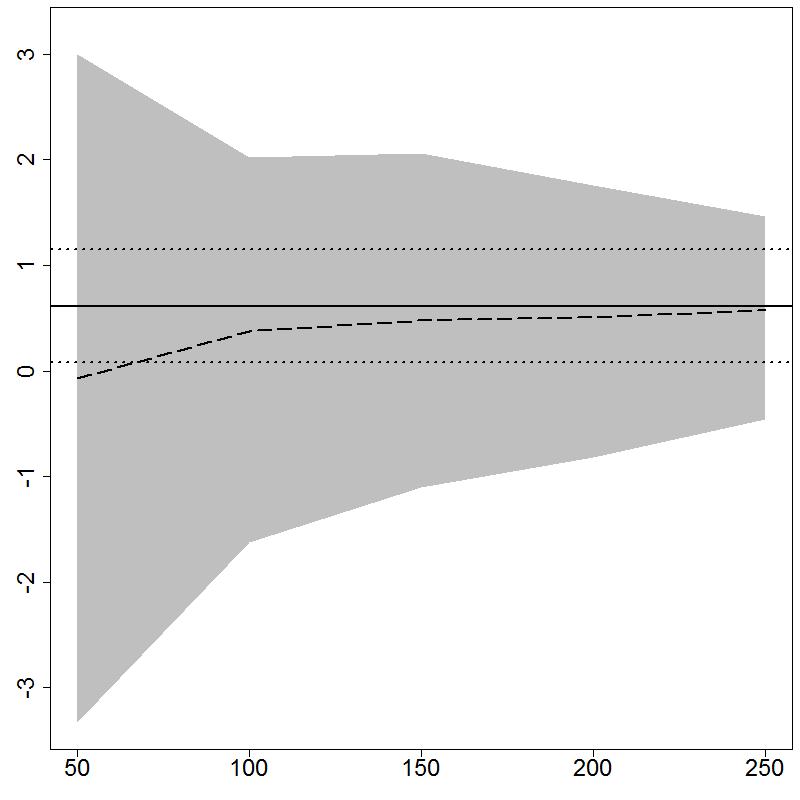}
\caption{Intercept ($\bgamma_\epsilon$)}
\end{subfigure}
\begin{subfigure}[]{0.48\textwidth}
\includegraphics[width=\textwidth]{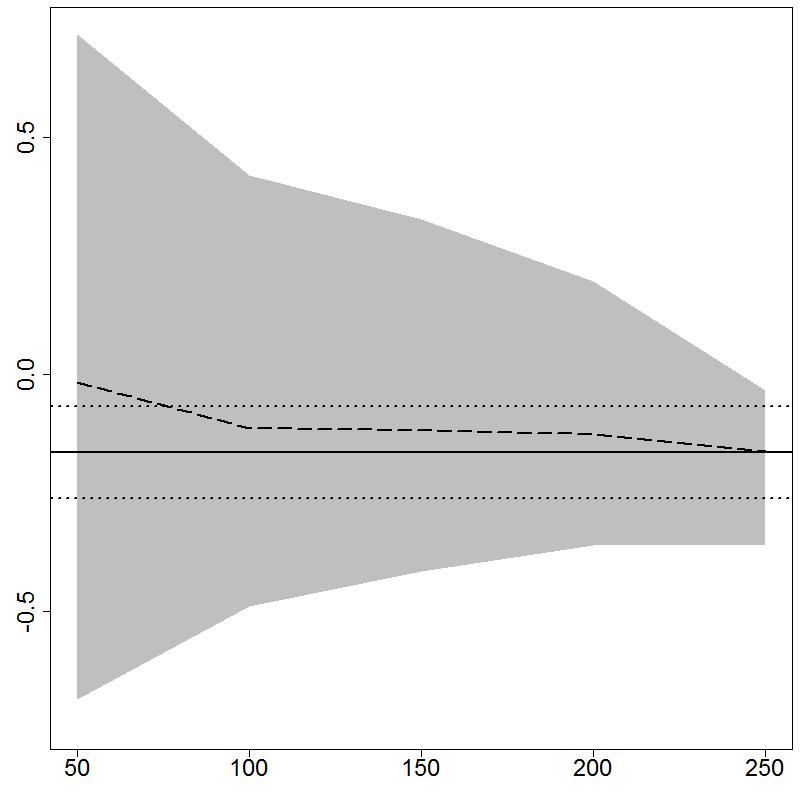}
\caption{(Q2.W) socialize ($\bgamma_\epsilon$)}
\end{subfigure} \\
\begin{subfigure}[]{0.48\textwidth}
\includegraphics[width=\textwidth]{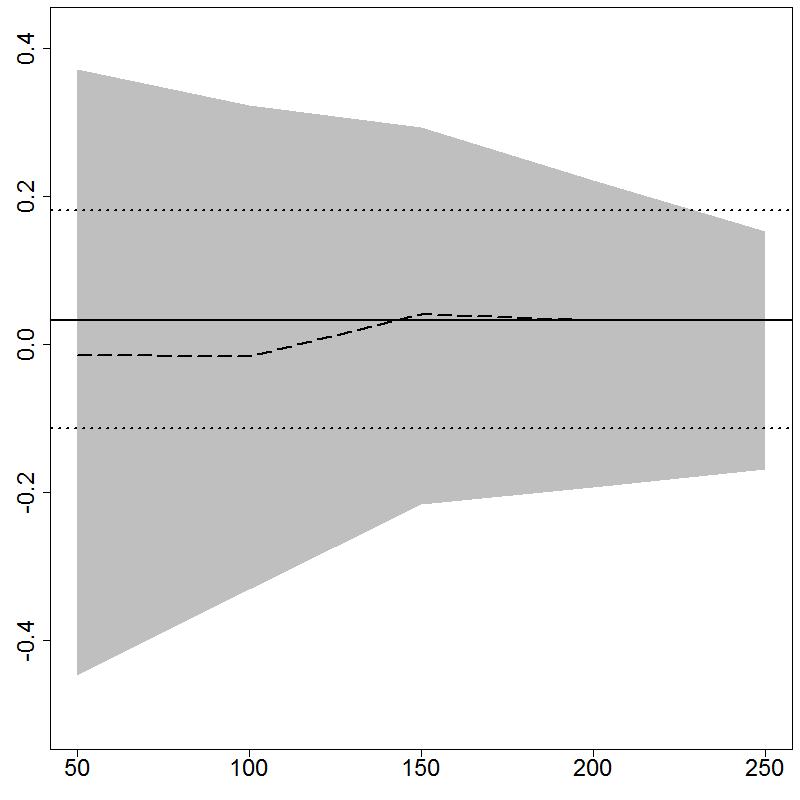}
\caption{(Q1.X) importance ($\bbeta_1$)}
\end{subfigure}
\begin{subfigure}[]{0.48\textwidth}
\includegraphics[width=\textwidth]{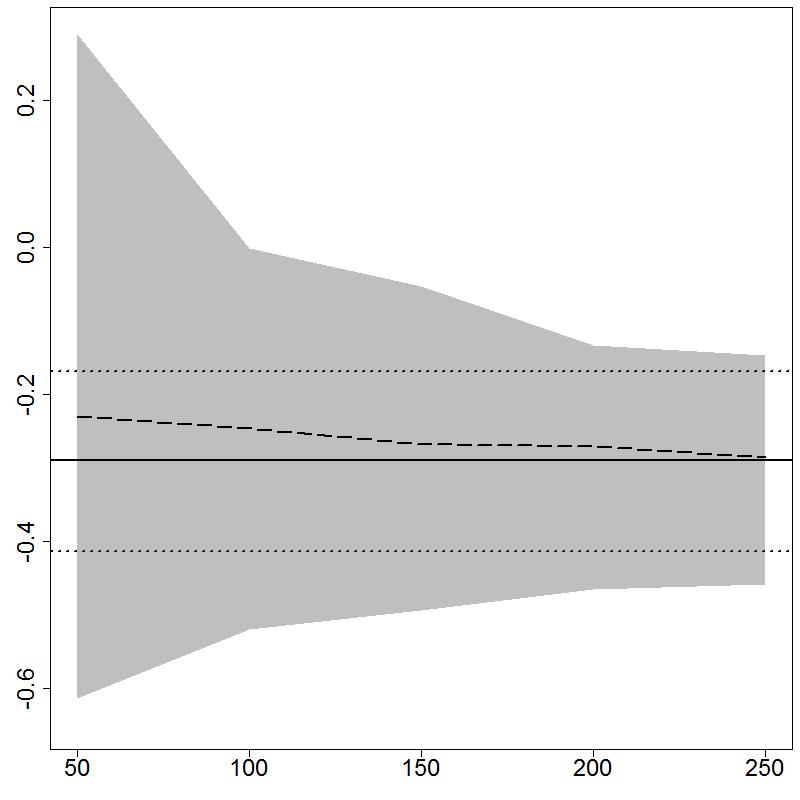}
\caption{(Q2.X) course subject ($\bbeta_1$)}
\end{subfigure}
\caption{Results for peer influence data running 100 egocentric subsamples for each value of $n_e\in\{50,100,150,200,250\}$.  The solid and dotted lines give respectively the posterior mean and 95\% CI bounds using the full network data.  The dashed line gives the average posterior mean over the 100 subsamples, and the shaded region corresponds to the 95\% CI bounds averaged over the 100 subsamples.}
\label{peerInflEgoFig1}
\end{figure}

Figure \ref{peerInflEgoFig1} shows the graphical results from the egocentric network analyses.  In each of the four plots we show for each $n_e\in\{50,100,150,200,250\}$ the average of the 100 posterior mean point estimates, the average of the 100 95\% credible interval lower and upper bounds, as well as the posterior mean and 95\% CI bounds when using the full network data.  From this we see that the results from the egocentric network are on average very similar to those results obtained from the full network, and in fact using 100 or more egos led to the average posterior mean being well within the 95\% full network CI for all parameters.  Since an egocentric network subsample will by necessity contain less information from the data, the 95\% CI bounds from the egocentric network must be larger than that obtained from the full network data;  that said, we see that the bounds shrink towards the full network bounds relatively quickly in most cases, and are quite close when we use 250 egos.

\begin{table}[h!]
\centering
\begin{tabular}{crrrr}
 & Estimate & SE & LB & UB \\ 
  \hline
Intercept ($\bgamma_\epsilon$) & 0.61* & 0.27 & 0.08 & 1.15 \\ 
(Q1.W) classmates ($\bgamma_\epsilon$) & 0.03 & 0.03 & -0.03 & 0.09 \\ 
(Q2.W) socialize ($\bgamma_\epsilon$) & -0.16* & 0.05 & -0.26 & -0.07 \\ 
(Q3.W) friends ($\bgamma_\epsilon$) & 0.01 & 0.06 & -0.10 & 0.12 \\ 
Inverse of network size ($\bgamma_\epsilon$)  & 0.12 & 0.14 & -0.15 & 0.39 \\ \hline
Intercept ($\bbeta_1$) & 3.04* & 0.24 & 2.58 & 3.50 \\ 
(Q1.X) importance ($\bbeta_1$)& 0.03 & 0.07 & -0.11 & 0.18 \\ 
(Q2.X) course subject ($\bbeta_1$)& -0.29* & 0.06 & -0.41 & -0.17 \\ \hline
$\sigma$ & 1.27 & 0.04  & 1.20 & 1.36
\end{tabular}
\caption{Results for peer influence data using the full data.  LB and UB correspond to the lower and upper bounds of the 95\% credible intervals respectively.  *'s indicate that the 95\% CI excluded zero.}
\label{peerInflResultsTab1}
\end{table}

\section{Discussion}
\label{conclusion}
\vspace{-1pc}
Network autocorrelation models have been invaluable in accounting for complex dependencies due to the subjects being connected through a network and in estimating the effect of the network via social influence on actor attributes of interest.  However, this class of models have all assumed uniform susceptibility to social influence.  We have proposed a non-linear hierarchical model which allows us to relax this strong assumption.  

Our proposed methods provide statistical tools to better understand social processes and to test social theories.  
For example, as mentioned in the introduction there are researchers who have posited that susceptibility to social influence varies based on gender, age, etc.; such claims can be tested by including these covariates into the design matrices corresponding to the susceptibilities (i.e., $W_x$ and $W_\epsilon$).  We have also discussed in Section \ref{rowNormalization} how row normalization of the adjacency matrix can be viewed as a subset of the proposed models by incorporating the reciprocal of individuals' local network sizes in $W_x$ or $W_\epsilon$.  In addition, by including the local clustering coefficient researchers may test how structural holes affect susceptibility \citep{burt2009structural}; by including betweenness one may investigate how one's position with regard to information flow affects susceptibility \citep{freeman1977set}; by including eigenvector centrality, one may investigate the effect of social capital on susceptibility \citep{bonacich1987power,borgatti1998network}.

When incorporating local network features into $W_x$ or $W_\epsilon$, one must be careful about highly correlated covariates and the subsequent negative inferential impacts.  Of course this is an important point of consideration in any linear model, but as local network features are often highly correlated with each other, such as local clustering coefficient and degree \citep{vazquez2002large}, researchers should be particularly cognizant of this issue.  Our simulation study did not seem to indicate that correlations as strong as $\approx0.6$ affected inference, but further study on this is merited

A drawback to the proposed methodology is the computational cost involved for large networks.  If one considers (\ref{generalPost}), it becomes apparent that one needs to compute the determinant and the inverse of an $n\times n$ matrix ($V$) and do so repeatedly during the optimization of $\pi(\bgamma_x,\bgamma_\epsilon | \by)$ and also for each draw $(\bgamma_x,\bgamma_\epsilon)$.  By using an egocentric subsample of the network data and capitalizing on Result \ref{result3}, researchers can investigate factors associated with susceptibility to peer influence without this computational burden, but clearly a better solution that makes use of all data is desired.  This is an important future direction for research.


\label{lastpage}

\bibliographystyle{asa}
\bibliography{heterogeneityInSocInflModels}

\end{document}